\documentclass[12pt]{article}
\usepackage{epsf}
\usepackage{amsfonts}
\usepackage{amsmath}
\usepackage[latin2]{inputenc}
\usepackage[T1]{fontenc}
\usepackage{epsfig}

\newlength{\dinwidth}
\newlength{\dinmargin}
\DeclareMathAlphabet{\scr}{U}{rsfs}{m}{n}
\begin{document}
\pagestyle{empty}
\begin{flushright}
IFT-2003-36
\end{flushright}
\vskip 2cm
\begin{center}
{\Huge
Singular gauge transformations and supersymmetry breakdown on warped
orbifolds}
\vspace*{1cm}
\end{center}
\vspace*{5mm} \noindent
\vskip 0.5cm
\centerline{\bf Zygmunt Lalak and Rados\l aw Matyszkiewicz}
\vskip 1cm
\centerline{\em Institute of Theoretical Physics}
\centerline{\em University of Warsaw, Poland}
\vskip 2cm
\vskip1cm
\centerline{\bf Abstract}
\vskip 0.5cm We have analyzed the breakdown of global supersymmetry by
a non-vanishing expectation value of the fifth component of the
graviphoton on warped $S^1 / Z_2$ orbifolds. It has been demonstrated
that the setups where such a breakdown is possible correspond to the
models where the true gauge symmetry on the orbifold, respecting the
$Z_2$--parities and periodicity, is broken by boundary terms. In the
tuned models, giving Randall--Sundrum warp factor, gauge symmetry
stays intact, and any $\langle A_5 \rangle$ can be gauged away without
violating supersymmetry.  \vskip3cm
\begin{flushleft}October 2003\end{flushleft}
\newpage
\pagestyle{plain}
\section{Introduction}
Brane-bulk supergravities generalize the concept of supersymmetry to
setups combining degrees of freedom  that propagate in subspaces of various dimensionalities, sometimes spatially disconnected \cite{Altendorfer:2000rr}--\cite{Lukas:1998yy}.
They provide  consistent field--theoretic framework
for the discussion of important aspects of modern string theory physics.
Most importantly, with supergravities that include lower dimensional branes it
is possible to study reliably supersymmetry breakdown and its transmission between various sectors of a model. In this note we shall
discuss a class of warped (gauged) supergravities with a non-zero background of the fifth component of the graviphoton switched on.
It turns out that such a background typically breaks supersymmetry completely
leading to four--dimensional $AdS_4$ models with no supersymmetric vacua.
The exceptional case is that of the models with tuned brane tensions  corresponding
to the Randall--Sundrum warped geometry. There the $\langle A_5 \rangle$
doesn't break supersymmetry.
In the present paper we discuss several examples with various
sets of boundary conditions, in particular the ones that correspond to super--bigravities. We also discuss in detail and clarify the relation between the original $U(1)$ gauge invariance that has been instrumental in the construction of the warped
supergravities, and the breaking of supersymmetry by non-zero $\langle A_5 \rangle$. It turns out that often in order to supersymmetrize the brane--bulk
Lagrangian one needs to sacrifice the original gauge invariance. These are precisely the cases where it is possible to break supersymmetry assuming non-vanishing $\langle A_5 \rangle$. One can think of such configurations as of the `would-be' Wilson lines of the broken gauge symmetry. Supersymmetry breakdown due to boundary conditions has been addressed by a number of papers \cite{Brax:2001xf}--\cite{Gersdorff:2003rq}.
In fact, the interesting point is that eventually the origin of
supersymmetry violation induced by non-zero $\langle A_5 \rangle$ can be traced back to fermionic boundary conditions
given by boundary mass terms of gravitini. The boundary masses are supersymmetry singlets, whereas boundary superpotentials, or boundary gaugino condensates,
which have been identified earlier as sources of supersymmetry violation,
do transform under supersymmetry.

\section{Setup}
To begin with, let us briefly summarize the brane--bulk
super--bigravity Lagrangian, constructed in \cite{Lalak:2003fu}. The simple N=2 d=5 supergravity multiplet contains metric tensor (represented by
the vielbein $e^m_\alpha$), two gravitini $\Psi^A_\alpha$ and one vector field $A_\alpha$ -- the graviphoton. We shall consider gauging of a $U(1)$ subgroup of the global $SU(2)_R$ symmetry of the 5d bulk Lagrangian. One adds to the initial bulk Lagrangian boundary terms that include brane tensions and/or gravitini mass terms on each brane.
The 5d action describing such a setup reads
        $S=\int_{M_5}\  ({\cal L}_{bulk}+{\cal L}_{brane})$,
         where
        \begin{eqnarray}
        &e_{5}^{-1}{\cal L}_{bulk}=&\frac{1}{2}R-\frac{3}{4}{\cal F}_{\alpha\beta}{\cal F}^{\alpha\beta}-\frac{1}{2\sqrt{2}}A_\alpha{\cal F}_{\beta\gamma}{\cal F}_{\delta\epsilon}\epsilon^{\alpha\beta\gamma\delta\epsilon}\nonumber\\&&-\frac{1}{2}\bar{\Psi}^A_\alpha\gamma^{\alpha\beta\gamma}D_\beta\Psi_{\gamma A}+\frac{3{\rm i}}{8\sqrt{2}}\left(\bar{\Psi}^A_\gamma\gamma^{\alpha\beta\gamma\delta}\Psi_{\delta A}+2\bar{\Psi}^{\alpha A}\Psi_{A}^{\beta}\right){\cal F}_{\alpha\beta}\nonumber\\&&-\frac{{\rm i}}{\sqrt{2}} {\cal P}_{AB}\bar{\Psi}^A_\alpha\gamma^{\alpha\beta}\Psi_{\beta}^B-\frac{8}{3} {\rm Tr}({\cal P}^{2})
        \end{eqnarray}
    and
    \begin{equation}
        {\cal L}_{brane} = \sum_i e_4\delta(y-y_i)\left(-\lambda_i- \bar{\Psi}_\mu^A\gamma^{\mu\nu}(M_i+\gamma_5\bar{M_i})_{A}^{\;B} \Psi_{\nu B}\right) \ .
   \end{equation}
        The $M_i$, $\bar{M_i}$ are constant matrices, symmetric in the symplectic indices, that denote gravitini mass terms on the brane at the fixed point $y_i$. The covariant derivative contains both gravitational and gauge connections:
        \begin{equation}
        D_\alpha\Psi_\beta^A=\nabla_\alpha\Psi_\beta^A+ A_\alpha{\cal P}^A_{\;B}\Psi_\beta^B\ ,
        \end{equation}
where ${\cal P}={\cal P}_a\,{\rm i}\sigma^a$ is the gauge prepotential. The pair of gravitini satisfies symplectic Majorana\- condition
 $\bar{\Psi}^A\equiv\Psi_A^\dagger\gamma_0=(\epsilon^{AB}\Psi_B)^TC$ where $C$ is the charge conjugation matrix and $\epsilon^{AB}$ is the antisymmetric $SU(2)_R$ metric (we use the convention $\epsilon_{12}=\epsilon^{12}=1$).
Supersymmetry transformations include singular terms proportional to the delta functions
        \begin{eqnarray}
        &&\delta e^m_\alpha=\frac{1}{2}\bar{\eta}^A\gamma^m\Psi_{\alpha A}\ ,\;\;\delta A_\alpha=-\frac{{\rm i}}{2\sqrt{2}}\bar{\Psi}_{\alpha}^A\eta_A\ ,\\&&\delta\Psi_{\alpha}^A=D_\alpha\eta^A- \frac{{\rm i}}{4\sqrt{2}}\left(\gamma_\alpha^{\;\beta\gamma}-4\delta_\alpha^{\;\beta}\gamma^\gamma\right){\cal F}_{\beta\gamma}\eta^A+\frac{\sqrt{2}{\rm i}}{3}{\cal P}^{AB}\gamma_\alpha\eta_B\ \nonumber\\&&\qquad\; + \epsilon^{-1}(y)\delta_\alpha^{\;5}\sum_ia_i\delta(y-y_i)(Q_i-\gamma_5\delta)^{A}_{\;B}\gamma_5\eta^B,
        \end{eqnarray}
    where $a_i= 1$ if the orbifold step function $\epsilon(y)$ `jumps up' at the fixed point $y_i$, or $a_i= -1$ if it `jumps down'. The $Q_i$ denotes ${\bf Z}_2$ parity operator acting locally on the gravitini sector\footnote{The parameters $\eta^A$ of the supersymmetry transformations obey the same boundary conditions as the 4d components of gravitini.} as follows:
    \begin{equation} \label{gbcond}
        \Psi^A_\mu(y_i-y)=\gamma_5(Q_i)^A_{\;B}\Psi^B_\mu(y_i+y)\ ,\quad\Psi^A_5(y_i-y)=-\gamma_5(Q_i)^A_{\;B}\Psi^B_5(y_i+y)\ .
    \end{equation}
The symplectic Majorana condition
and the normalization $(Q_i)^2=1$ imply $Q_i=(q_i)_a \sigma^a$, where $(q_i)_a$ are real parameters.

In the general case \cite{Brax:2001xf} one can write down the prepotential
as follows:
     $ {\cal P} = g_R \epsilon(y)  {\cal R} + g_{S}  {\cal S}$,
     where $ {\cal R}=r_a\, {\rm i} \sigma^a$ commutes and  $ {\cal S}=s_a\, {\rm i}\sigma^a$ anticommutes with each $Q_i$.

     The closure of the supersymmetry algebra provides relations  between parameters of the
boundary Lagrangian and the prepotential:
  \begin{eqnarray} \label{rownanienaznikanie}
    0=&\delta(y-y_i)\bar{\Psi}_\mu^A\gamma^\mu\bigg[&(M_i-\gamma_5\bar{M}_i)_{A}^{\;B} {\cal P}_{B}^{\;C}+a_i\frac{1}{2}\epsilon^{-1}(y) {\cal P}_{A}^{\;B}(Q_i+\gamma_5\delta)_{B}^{\;C}+a_ig_R\gamma_5  {\cal R}_{A}^{\;C}\nonumber\\&&+\frac{{\rm i}}{4\sqrt{2}}\lambda_i\delta_{A}^{\;C}+a_i\frac{{\rm i}}{4\sqrt{2}}\lambda_i\epsilon(y)\left(\gamma_5(M_i)_{A}^{\;C}-(\bar{M}_i)_{A}^{\;C}\right)\nonumber\\&&+\frac{{\rm i}}{4\sqrt{2}}\lambda_i\left(\frac{1}{2}\gamma_5(Q_i)_{A}^{\;C}+\frac{1}{2}\delta_{A}^{\;C}\right)\bigg]\eta_C\ .
  \end{eqnarray}

Let us assume the prepotential of the form $ {\cal P}_A^{\;B}=g{\rm i}(\sigma_1)_A^{\;B}$ and $(Q_0)_A^{\;B}=(Q_\pi)_A^{\;B}=(\sigma_3)_A^{\;B}$. Let us allow only the even components of gravitini to have mass terms on the branes
     \begin{equation}
     (M_{0,\pi})_A^{\;B}=\frac{1}{2}\alpha_{0,\pi}(\sigma_1)_A^{\;B}\ ,\quad (\bar{M}_{0,\pi})_A^{\;B}=\frac{1}{2}{\rm i}\alpha_{0,\pi}(\sigma_2)_A^{\;B}\ .
       \end{equation}
     Then the boundary conditions take the form
         \begin{eqnarray}
    &&\epsilon^{-1}(y) \delta(y) \gamma_5(\eta_-)^A=-\delta(y)\alpha_{0}(\sigma_1)^A_{\;B}(\eta_+)^B\ ,\nonumber\\  &&\epsilon^{-1}(y) \delta(y-\pi r_c) \gamma_5(\eta_-)^A=\delta(y-\pi r_c)\alpha_{\pi}(\sigma_1)^A_{\;B}(\eta_+)^B\label{warbrzadsbigravit1}\ ,
      \end{eqnarray}
     where we have decomposed fermions into the even $(+)$ and odd $(-)$
 components
    \begin{equation} \label{defpmspinors}
     (\eta_{\pm})^A=\frac{1}{2}(\delta\pm\gamma_5\sigma_{3})^{A}_{\;B}\eta^B\ .
    \end{equation}
     The equations (\ref{rownanienaznikanie}) are satisfied, if
       \begin{equation}
 \lambda_0=-g4\sqrt{2}\frac{2\alpha_{0}}{1+\alpha_{0}^{2}}\ ,\qquad  \lambda_\pi=-g4\sqrt{2}\frac{2\alpha_{\pi}}{1+\alpha_{\pi}^{2}}\ .
       \end{equation}
For $g=\frac{3}{4}\sqrt{2}k$ the bosonic part of the Lagrangian reads
       \begin{equation} \label{lagrs}
    S=\int d^5 x \sqrt{-g_5} (\frac{1}{2}R + 6 k^2)-  6 \int d^5 x\sqrt{-g_4}k  (T_0\delta(y) +T_\pi \delta(y-\pi r_c))\ ,
       \end{equation}
       where $T_{0,\pi}=-2\alpha_{0,\pi}/(1+\alpha_{0,\pi}^{2})$. Note that $|T_{0,\pi}|\leq 1$. For $T_{0}=-T_{\pi}=1$ we obtain supersymmetric Randall-Sundrum model, while for other values of $T_{0,\pi}$ we have $AdS_5$ in the bulk with $AdS_4$ foliations (the super--bigravity for example).

       \section{Supersymmetry breakdown}\label{secsusybreak}
  Let us assume a nonzero expectation value of $A_{5}$. Let us solve Killing equation to check whether supersymmetry remains unbroken:
       \begin{equation}
    D_\alpha\eta^A-\frac{\sqrt{2}{\rm i}}{3}{\cal P}^{A}_{\;B}\gamma_\alpha\eta^B-\left(\delta(y)-\delta(y-\pi r_c)\right)\epsilon^{-1}(y)\delta_\alpha^5(\delta-\sigma_3\gamma_5)^{A}_{\;B}\eta^B=0\ .
       \end{equation}
       For the RS background ($\alpha_{0}=-\alpha_{\pi}=-1$) we can write
       \begin{eqnarray}
        &&0=\partial_\mu\eta_\pm^A-\frac{1}{2}k\epsilon(y)\gamma_\mu\gamma_5\eta_\mp^A+\frac{1}{2}k(\sigma_1)^A_{\;B}\gamma_\mu\eta_\pm^B\ ,\label{killingbag4}\\&&0=\partial_5\eta_+^A+{\rm i}e(\sigma_1)^A_{\;B}\eta_-^B+\frac{1}{2}k(\sigma_1)^A_{\;B}\gamma_5\eta_-^B\ ,\label{killingbag5p}\\&&0=\partial_5\eta_-^A+{\rm i}e(\sigma_1)^A_{\;B}\eta_+^B+\frac{1}{2}k(\sigma_1)^A_{\;B}\gamma_5\eta_+^B-2(\delta(y)-\delta(y-\pi r_c))\epsilon^{-1}(y)\eta_-^A\ ,\label{killingbag5m}
    \end{eqnarray}
       where $e=g\langle A_{5}\rangle$.
    The equation (\ref{killingbag4}) is satisfied by $\eta_-^A=\epsilon(y)\gamma_5(\sigma_1)^A_{\;B}\eta_+^B$, where we have assumed that the Killing spinor doesn't
 depend on $x_\mu$.
       One can easily find the solution of (\ref{killingbag5p}) and (\ref{killingbag5m})
       \begin{eqnarray} \label{solutkillingspin}
       &&\eta_+^1=e^{-\frac{1}{2}(k+2ie)|y|}\hat{\eta}_R\ , \qquad \;\:\,\eta_-^1=\epsilon(y)e^{-\frac{1}{2}(k-2ie)|y|}\hat{\eta}_L\ ,\nonumber\\&&\eta_+^2=-e^{-\frac{1}{2}(k-2ie)|y|}\hat{\eta}_L \ ,\qquad\eta_-^2=\epsilon(y)e^{-\frac{1}{2}(k+2ie)|y|}\hat{\eta}_R  \ ,
       \end{eqnarray}
       where $\hat{\eta}$ is a four--dimensional Majorana spinor in flat space. Notice, that first and second components of the Killing spinors have phases, which are complex conjugates of each other. In fact, this relation is required by the 5d Majorana condition. Thus there exists a global unbroken supersymmetry that gives rise to a flat N=1 supergravity in 4d with susy preserving vacua.

       Let us now turn to the detuned case. As an example we consider the super--bigravity. Taking
      \begin{equation} \label{alphamasy}
    \alpha_0=-\frac{\cosh(k\pi r_{c}/2)\pm 1}{\sinh(k\pi r_{c}/2)}\ ,\quad  \alpha_\pi=-\frac{\cosh(k\pi r_{c}/2)\pm 1}{\sinh(k\pi r_{c}/2)}
      \end{equation}
      we obtain the bosonic action of the 4d bigravity
       \begin{equation} \label{lagads}
    S= \int d^5 x \sqrt{-g_5} (\frac{1}{2}R + 6 k^2)-  6 \int d^5 x\sqrt{-g_4}k T (\delta(y) +\delta(y-\pi r_c))\ ,
       \end{equation}
       where $k=\frac{2\sqrt{2}}{3}g$ and $T=\tanh(k\pi r_{c}/2)$.
       Gravitational background does not admit a flat 4d Minkowski foliation, and the consistent
solution is that of $AdS_4$ branes:
        \begin{equation}
        ds^{2}=a^{2}(y)\bar{g}_{\mu\nu}dx^{\mu}dx^{\nu}+dy^{2}\ ,
        \end{equation}
        where
        \begin{equation}
        a(y)=\frac{\sqrt{-\bar{\Lambda}}}{k}\cosh\left(k|y|-\frac{k\pi r_{c}}{2}\right)\ ,
        \end{equation}
        and $\bar{g}_{\mu\nu}dx^{\mu}dx^{\nu}=\exp(-2\sqrt{-\bar{\Lambda}}x_{3})(-dt^{2}+dx^{2}_{1}+dx_{2}^{2})+dx_{3}^{2}$ is the four--dimensional $AdS$ metric.
The radius of the fifth dimension is determined in terms of the brane tensions
        \begin{equation}
        k\pi r_{c}=\ln\left(\frac{1+T}{1-T}\right)\ .
        \end{equation}
        Normalization $a(0)=1$ leads to the fine--tuning relation $\bar{\Lambda}=(T^{2}-1)k^{2}<0$.

 Notice in (\ref{alphamasy}), that we have two possibilities for the brane
gravitini masses: $\alpha_0=1/\alpha_\pi$ and $\alpha_0=\alpha_\pi$. In the first case five--dimensional vacuum spontaneously breaks all supersymmetries, while in the second N=1 supersymmetry is preserved. To justify this observation, let us assume
$\alpha_0=\alpha_\pi=-\alpha$, where
        \begin{equation}
    \alpha=\frac{\cosh(k\pi r_{c}/2)- 1}{\sinh(k\pi r_{c}/2)}\ .
      \end{equation}
     Killing equations in the presence of a non-zero expectation value of the $\langle A_5\rangle=e/g$, take the form
     \begin{eqnarray}
        &&0=\bar{\nabla}_\mu\eta_\pm^A+\frac{1}{2}k\epsilon(y)\tanh\left(k|y|-\frac{k\pi r_{c}}{2}\right)\gamma_\mu\gamma_5\eta_\mp^A+\frac{1}{2}k(\sigma_1)^A_{\;B}\gamma_\mu\eta_\pm^B\ ,\label{killingbag4bigrav}\\&&0=\partial_5\eta_+^A+{\rm i}e(\sigma_1)^A_{\;B}\eta_-^B+\frac{1}{2}k(\sigma_1)^A_{\;B}\gamma_5\eta_-^B\ ,\label{killingbag5bigrav}\\&&0=\partial_5\eta_-^A+{\rm i}e(\sigma_1)^A_{\;B}\eta_+^B+\frac{1}{2}k(\sigma_1)^A_{\;B}\gamma_5\eta_+^B-2(\delta(y)-\delta(y-\pi r_c))\epsilon^{-1}(y)\eta_-^A\ ,\label{killingbag5mbigrav}
    \end{eqnarray}
       where $\bar{\nabla}_\mu$ denotes covariant derivative with respect to the four--dimensional $AdS$ geometry. Note that the equations have imaginary coefficients, hence we have to consider complex amplitudes in the solution. The appropriate decomposition of the Killing spinors reads
       \begin{equation}
       \eta_+^A=\left(\begin{array}{c}\phi_+(y)\hat{\eta}_R\\-\phi_+^\star(y)\hat{\eta}_L\end{array}\right)^A\ ,\qquad\eta_-^A=\epsilon(y)\left(\begin{array}{c}\phi_-^\star(y)\hat{\eta}_L\\\phi_-(y)\hat{\eta}_R\end{array}\right)^A\ ,
       \end{equation}
       where $\hat{\eta}$ denotes the Killing spinor in the $AdS_4$ which satisfies:
      $
       \left(\bar{\nabla}_\mu-\frac{1}{2}\sqrt{-\bar{\Lambda}}\hat{\gamma}_\mu\right)\hat{\eta}=0\ .
      $

The equations (\ref{killingbag4bigrav}),(\ref{killingbag5bigrav}) and (\ref{killingbag5mbigrav})  can be solved by
       \begin{eqnarray}
     \phi_+(y)=&N\cos\left(e|y|\right)\cosh\left(\frac{k|y|}{2}-\frac{k\pi r_{c}}{4}\right)+{\rm i}N\sin\left(e|y|\right)\sinh\left(\frac{k|y|}{2}-\frac{k\pi r_{c}}{4}\right)&\ ,\nonumber\\\phi_-(y)=&-N\cos\left(e|y|\right)\sinh\left(\frac{k|y|}{2}-\frac{k\pi r_{c}}{4}\right)-{\rm i}N\sin\left(e|y|\right)\cosh\left(\frac{k|y|}{2}-\frac{k\pi r_{c}}{4}\right)&\ ,
       \end{eqnarray}
        where $N$ is a normalization constant.
      One can easily check that boundary conditions (\ref{warbrzadsbigravit1}) are satisfied only if
      \begin{equation}
      \frac{\sin\left(e\pi r_{c}\right)}{\cosh\left(k\pi r_{c}/4\right)}=0\ ,
      \end{equation}
       which means that the $\langle A_5\rangle$ background does not break supersymmetry for $e= p/r_c$, where $p$ is an integer ($ p\in {\bf Z}$), or in the limit $r_c\rightarrow\infty$, which implies $\lambda_{0/\pi}\rightarrow 6k$.

       \section{Singular gauge transformations}
       Typically, especially in models considered on ${\bf S^{1}}$, one can break supersymmetry by a non-zero expectation value of $A_5$. To see why the situation is more complicated  on an orbifold  ${\bf S^{1}}/{\bf Z_{2}}$, let us attempt to go to the picture where $\langle A_5\rangle =0$. We can try to do this using gauge transformation of the form, \cite{Lalak:2002kx},
       \begin{equation}
       \Psi_M\longrightarrow  e^{{\cal P}\Omega(y)}\Psi_M\ ,\quad \eta\longrightarrow  e^{{\cal P}\Omega(y)}\eta \ ,
       \end{equation}
       with ${\cal P}_A^{\;B}=g{\rm i}(\sigma_1)_A^{\;B}$, where we consider the most general form of the transformation parameter $\Omega(y)=\omega(|y|)\epsilon(y)$.
Note, that in this paper by a gauge transformation we mean exclusively
transformations which are `legal' on the orbifold, i.e. the ones which are periodic and preserve $Z_2$--parities of all fields.
We will find the explicit form of the $\Omega(y)$ requiring that $\delta A_5=-e/g$ and gauge invariance of the bulk plus brane action.

        To be consistent with the `epsilon rules' (see \cite{Lalak:2003fu}), we have to require that gauge transformations do not change the powers of the $\epsilon(y)$
    \begin{eqnarray}
    &&\Psi_A^+\longrightarrow \Psi_A^+ + {\rm i}g\omega\epsilon^{-1}(\sigma_1)_{A}^{\;B}\Psi_B^-\ ,\nonumber\\&&\Psi_A^-\longrightarrow \Psi_A^- + {\rm i}g\omega\epsilon(\sigma_1)_{A}^{\;B}\Psi_B^+\ .
    \end{eqnarray}
    One can write the finite transformations as follows
    \begin{eqnarray} \label{finitetransfgauge}
       \Psi^\pm_A\longrightarrow&&\frac{1}{2}\left(\delta\pm\gamma_5\sigma_3\right) e^{{\rm i}g\sigma_1\omega\epsilon}\Psi_A =\sum_{n=0}^\infty\frac{({\rm i}g\omega)^{2n}}{(2n)!}\Psi^\pm_A+\sum_{n=0}^\infty\frac{({\rm i}g\omega)^{2n+1}}{(2n+1)!}\epsilon^{\mp 1}\sigma_1\Psi^\mp_A\nonumber\\&&=\cos\left(g\omega\right)\Psi^\pm_A+{\rm i}\epsilon^{\mp 1}\sin\left(g\omega\right)\sigma_1\Psi^\mp_A\ .
       \end{eqnarray}
       Notice, that in this case we should treat $e^{{\rm i}g\sigma_1\omega\epsilon}$ not as a function of $\epsilon$, but rather as a symbolic shorthand for the expression (\ref{finitetransfgauge}). If we restrict ourselves to the bulk Lagrangian, both forms coincide. The difference becomes important for the singular terms proportional to the $\delta(y)$ function.

       The Lagrangian variation under gauge transformations includes:
       \begin{eqnarray}
        &&\frac{1}{2}(\bar{\Psi}_\pm)^A_\mu\gamma^{\mu\nu}\gamma^5 \partial_5(\Psi_\mp)_{\nu A}+\frac{1}{2}(\bar{\Psi}_\pm)^A_\mu\gamma^{\mu\nu}\gamma^5 A_5 {\rm i}g(\sigma_1)_{A}^{\;B}(\Psi_\pm)_{\nu B}\nonumber\\&&-(\bar{\Psi}_+)^A_\mu\gamma^{\mu\nu}(\sigma_1)_{A}^{\;B}(\Psi_+)_{\nu B}\left(\alpha_{0}\delta(y)+\alpha_{\pi}\delta(y-\pi r_c)\right)\longrightarrow\nonumber\\&&\longrightarrow \frac{1}{2}(\bar{\Psi}_\pm)^A_\mu\gamma^{\mu\nu}\gamma^5 \partial_5(\Psi_\mp)_{\nu A}+\frac{1}{2}(\bar{\Psi}_\pm)^A_\mu\gamma^{\mu\nu}\gamma^5 {\rm i}g\omega'\epsilon(\sigma_1)_A^{\;B}(\Psi_\pm)_{\nu B}\nonumber\\&&\quad\,\pm
{\rm i}\cos\left(g\omega\right)\sin\left(g\omega\right)
(\bar{\Psi}_\pm)^A_\mu \gamma^{\mu\nu}\gamma^5 \sigma_1(\Psi_\pm)_{\nu A}\epsilon^{-1}\epsilon^{\pm 1}\left(\delta(y)-\delta(y-\pi r_c)\right)\nonumber\\&&\quad\,\pm\sin^{2}\left(g\omega\right)
(\bar{\Psi}_\mp)^A_\mu \gamma^{\mu\nu}\gamma^5 (\Psi_\pm)_{\nu A}\epsilon^{-1}\left(\delta(y)-\delta(y-\pi r_c)\right)\nonumber\\&&\quad\,+\frac{1}{2}(\bar{\Psi}_\pm)^A_\mu\gamma^{\mu\nu}\gamma^5 A_5  {\rm i}g(\sigma_1)_A^{\;B}(\Psi_\pm)_{\nu B}-\frac{1}{2}(\bar{\Psi}_\pm)^A_\mu\gamma^{\mu\nu}\gamma^5 {\rm i} e(\sigma_1)_A^{\;B}(\Psi_\pm)_{\nu B}\nonumber\\&&\quad\,-\cos^{2}\left(g\omega\right)(\bar{\Psi}_+)^A_\mu\gamma^{\mu\nu}(\sigma_1)_{A}^{\;B}(\Psi_+)_{\nu B}\left(\alpha_{0}\delta(y)+\alpha_{\pi}\delta(y-\pi r_c)\right)\nonumber\\&&\quad\,-\sin^{2}\left(g\omega\right)(\bar{\Psi}_-)^A_\mu\gamma^{\mu\nu}(\sigma_1)_{A}^{\;B}(\Psi_-)_{\nu B}\epsilon^{-2}\left(\alpha_{0}\delta(y)+\alpha_{\pi}\delta(y-\pi r_c)\right)\nonumber\\&&\quad\,\mp
{\rm i}\cos\left(g\omega\right)\sin\left(g\omega\right)
(\bar{\Psi}_\pm)^A_\mu \gamma^{\mu\nu}(\Psi_\mp)_{\nu A}\epsilon^{-1}\left(\alpha_{0}\delta(y)+\alpha_{\pi}\delta(y-\pi r_c)\right)\ .
       \end{eqnarray}
       Gauge invariance in the bulk requires
       \begin{equation} \label{transfgaugea5}
       \omega'(y)=\frac{e}{g}\epsilon(y)\Longrightarrow \omega(y)=\frac{e}{g}|y|\ .
       \end{equation}
       One can verify, that such a transformation removes imaginary phases in the solution (\ref{solutkillingspin}).

       It is worth noticing that in terms of the odd parameter $\Omega$, the gauge transformation of the $A_5$ reads
       \begin{equation}
       A_5 \longrightarrow A_5-\epsilon\partial_5 \left(\Omega \epsilon^{-1}\right)\ .
       \end{equation}

       One can check that on the branes the Lagrangian is not gauge invariant for $|\alpha_{0,\pi}|\neq 1$, and the uncancelled variation reads
       \begin{eqnarray}
       \delta{\cal L}=&&{\rm i}\cos\left(g\omega\right)\sin\left(g\omega\right)
(\bar{\Psi}_+)^A_\mu \gamma^{\mu\nu}\gamma^5 \sigma_1(\Psi_+)_{\nu A}\left((1-\alpha_{0}^{2})\delta(y)-(1-\alpha_{\pi}^{2})\delta(y-\pi r_c)\right)\ ,\nonumber\\&&+\sin^{2}\left(g\omega\right)
(\bar{\Psi}_-)^A_\mu \gamma^{\mu\nu}\gamma^5 (\Psi_+)_{\nu A}\epsilon^{-1}\left((1-\alpha_{0}^{2})\delta(y)-(1-\alpha_{\pi}^{2})\delta(y-\pi r_c)\right) \ .
       \end{eqnarray}
       The above variation vanishes for $\sin\left(g\omega(0)\right)=0$ and $\sin\left(g\omega(\pi r_c)\right)=0$. The same conclusion can be obtained in a different way. One can analyze the action of gauge transformations  on the boundary conditions (\ref{warbrzadsbigravit1}). These change under the gauge transformation
       \begin{equation} \label{transformacja}
       \eta^\pm_A\longrightarrow\cos\left(g\omega\right)\eta^\pm_A+{\rm i}\epsilon^{\mp 1}\sin\left(g\omega\right)\sigma_1\eta^\mp_A
       \end{equation}
       into
        \begin{eqnarray}
    &&\epsilon^{-1} \delta(y) \cos\left(g\omega\right)\gamma_5(\eta^-)^A+{\rm i}\delta(y)\sin\left(g\omega\right)(\sigma_1)^A_{\;B}\gamma_5(\eta^+)^B\nonumber\\&&\qquad\qquad=-\delta(y)\alpha_{0}\cos\left(g\omega\right)(\sigma_1)^A_{\;B}(\eta^+)^B-{\rm i}\epsilon^{-1}\delta(y)\alpha_{0}\sin\left(g\omega\right)(\eta^-)^A\ ,\label{warbrzadsbigravittran1}\nonumber\\   &&\epsilon^{-1} \delta(y-\pi r_{c}) \cos\left(g\omega\right)\gamma_5(\eta^-)^A+{\rm i}\delta(y-\pi r_{c})\sin\left(g\omega\right)(\sigma_1)^A_{\;B}\gamma_5(\eta^+)^B\nonumber\\&&\qquad\qquad=\delta(y-\pi r_{c})\alpha_{\pi}\cos\left(g\omega\right)(\sigma_1)^A_{\;B}(\eta^+)^B+{\rm i}\epsilon^{-1}\delta(y-\pi r_{c})\alpha_{\pi}\sin\left(g\omega\right)(\eta^-)^A\label{warbrzadsbigravittran2}\ .
      \end{eqnarray}
    These boundary conditions are invariant under the gauge transformation if $\sin\left(g\omega(0)\right)=0$ and $\sin\left(g\omega(\pi r_c)\right)=0$, or for $|\alpha_{0,\pi}|=1$. For the specific $\omega(y)$ given by (\ref{transfgaugea5}) we obtain quantization condition for the allowed backgrounds $e_p$, which singles out gauge transformations which do not change boundary conditions:
    \begin{equation} \label{kwantyzacja}
    e_p= p/r_c\ ,
    \end{equation}
    where $ p\in {\bf Z}$. We can parametrize different classes of the detuned models by a parameter
    $\theta\in \langle 0,\frac{1}{g r_c})$ such that for given $\alpha_{0,\pi}$ the vacuum
    expectation value of $A_5$ equals $\langle A_5\rangle =\theta+\frac{p}{g r_c}$, where $ p\in {\bf Z}$. Models belonging to the same class are connected to each other by transformations (\ref{transformacja}) with (\ref{transfgaugea5}) and (\ref{kwantyzacja}). For $\alpha_0=\alpha_\pi=-\alpha$ the class with unbroken N=1 supersymmetry is labelled by $\theta=0$. One can check that $\theta=\frac{1}{2 g r_c}$ corresponds to the flipped super--bigravity. The redefinition (an `illegal' gauge transformation)
    \begin{eqnarray}
      &&\eta^\pm_A\longrightarrow\cos\left(\frac{|y|}{2 r_{c}}\right)\eta^\pm_A+{\rm i}\epsilon^{\mp 1}\sin\left(\frac{|y|}{2 r_{c}}\right)\sigma_1\eta^\mp_A\ ,\\&&A_5\longrightarrow A_5-\frac{1}{2g r_c}
    \end{eqnarray}
    transforms this model to the frame, where $\alpha_0=1/\alpha_\pi=-\alpha$ with  $\langle A_5\rangle =0$. In the same way one can show that configuration $\alpha_0=1/\alpha_\pi=-\alpha$ with  $\langle A_5\rangle =\frac{1}{2 g r_c}$ corresponds, upon the same redefinition, to the configuration $\alpha_0=\alpha_\pi=-\alpha$ with vanishing $\langle A_5\rangle$, where N=1 supersymmetry is unbroken.

    It is worth noticing that even for a function $\omega$ which is independent of the fifth coordinate, the brane Lagrangian is not invariant under the gauge transformation as long as $|\alpha_{0}|\neq 1$ or $|\alpha_{\pi}|\neq 1$, in fact even the global $U(1)$ symmetry is explicitly broken. Supersymmetry breakdown by a non-zero vacuum expectation value of $A_5$ can be understood as a consequence of the explicit breaking of the gauged $U(1)$ symmetry. To be more explicit, let us assume that for some values of the parameters $\alpha_{0}$, $\alpha_{\pi}$ and $\langle A_5\rangle=e/g$, the N=1 supersymmetry  stays unbroken, and that there exists a Killing spinor which satisfies boundary conditions generated by $\alpha_{0}$ and $\alpha_{\pi}$. Let us imagine choosing another expectation value of $A_5$ ($\langle A_5\rangle=e'/g$) such that $e'-e\neq\frac{p}{r_c}$, where $ p\in {\bf Z}$. One can gauge-transform this model to the frame where $\langle A_5\rangle=e/g$. The bulk Lagrangian is invariant under such a transformation, hence the solution for the Killing spinor in the bulk remains unchanged. However, the brane Lagrangian is not invariant and one obtains different boundary conditions with $\alpha_{0}^{\prime}\neq \alpha_{0}$ or $\alpha_{\pi}^{\prime}\neq \alpha_{\pi}$ (`prime' denotes parameters after the transformation), which are not satisfied by the bulk solution and, consequently, supersymmetry must be broken. On the other hand, in the `tuned' case one can always gauge away the non-zero vacuum expectation value of $A_5$, hence, if supersymmetry is unbroken for some $\langle A_5\rangle$, this implies that it remains unbroken for any $\langle A_5\rangle$.

\section{Supersymmetric Randall-Sundrum model with ${\bf Z_{2}}$-odd prepotential}
It is interesting to check whether the same analysis can be repeated for the FLP model. Let us take the prepotential ${\cal P}_A^{\;B}=g{\rm i}\epsilon(y)(\sigma_3)_A^{\;B}$ and $(Q_0)_A^{\;B}=(Q_\pi)_A^{\;B}=(\sigma_3)_A^{\;B}$. We do not put any gravitini mass terms on the brane.
     The equations (\ref{rownanienaznikanie}) reduce to
       \begin{equation}
 \lambda_0=g4\sqrt{2}\ ,\qquad  \lambda_\pi=-g4\sqrt{2}\ .
       \end{equation}
For $g=\frac{3}{4}\sqrt{2}k$ we obtain the bosonic part of the Randall-Sundrum model.

  Let us assume nonzero expectation value of $A_{5}=e/g$. Killing equation in the RS background reads
       \begin{eqnarray}
        &&0=\partial_\mu\eta_\pm^A-\frac{1}{2}k\epsilon(y)\gamma_\mu\gamma_5\eta_\mp^A+\frac{1}{2}k\epsilon(y)(\sigma_3)^A_{\;B}\gamma_\mu\eta_\mp^B\ ,\label{killingflp4}\\&&0=\partial_5\eta_\pm^A+{\rm i}e\epsilon(y)(\sigma_3)^A_{\;B}\eta_\pm^B+\frac{1}{2}k\epsilon(y)(\sigma_3)^A_{\;B}\gamma_5\eta_\pm^B\ .\label{killingflp5}
    \end{eqnarray}

    The equation (\ref{killingflp4}) implies $\eta_-^A=0$, where it has been assumed that the Killing spinor doesn't
 depend on $x_\mu$.
       One can easily find the solution for $\eta_+^A$
       \begin{eqnarray} \label{solutkillingspinflp}
       &&\eta_+^1=e^{-\frac{1}{2}(k+2ie)|y|}\hat{\eta}_R\ , \qquad \;\:\,\eta_+^2=-e^{-\frac{1}{2}(k-2ie)|y|}\hat{\eta}_L\ ,
       \end{eqnarray}
       where $\hat{\eta}$ is a four--dimensional Majorana spinor in flat space.

       Let us turn to the picture where $\langle A_5\rangle =0$. We can do this using the gauge transformation
       \begin{equation}
       \Psi_M\longrightarrow  e^{{\cal P}\Omega(y)}\Psi_M\ ,\quad \eta\longrightarrow  e^{{\cal P}\Omega(y)}\eta \ .
       \end{equation}
       In this case we do not face problems with the singular terms, because we do not consider gravitini mass terms on the brane and  $\partial_5 (e^{{\cal P}\Omega(y)})$ does not produce $\delta(y)$ function (${\cal P}\Omega(y)$ is even). So, we can simply write finite transformation as follows:
       \begin{equation} \label{finitetransfgaugeflp}
       \Psi^\pm_A\longrightarrow e^{{\rm i}\epsilon g\sigma_3\Omega}\Psi^\pm_A \ ,\qquad A_5\longrightarrow A_5 -\frac{e}{g}\ .
       \end{equation}
        The Lagrangian variation under the $U(1)$ gauge transformation includes:
       \begin{eqnarray}
        &&\frac{1}{2}(\bar{\Psi}_\pm)^A_\mu\gamma^{\mu\nu}\gamma^5 \partial_5(\Psi_\mp)_{\nu A}+\frac{1}{2}(\bar{\Psi}_\pm)^A_\mu\gamma^{\mu\nu}\gamma^5 A_5 {\rm i}\epsilon g(\sigma_3)_{A}^{\;B}(\Psi_\mp)_{\nu B}\longrightarrow \nonumber\\&&\longrightarrow \frac{1}{2}(\bar{\Psi}_\pm)^A_\mu\gamma^{\mu\nu}\gamma^5 \partial_5(\Psi_\mp)_{\nu A}+\frac{1}{2}(\bar{\Psi}_\pm)^A_\mu\gamma^{\mu\nu}\gamma^5 {\rm i} g(\epsilon \Omega)'(\sigma_3)_A^{\;B}(\Psi_\mp)_{\nu B}\nonumber\\&&\quad\,+\frac{1}{2}(\bar{\Psi}_\pm)^A_\mu\gamma^{\mu\nu}\gamma^5 A_5 {\rm i}\epsilon g(\sigma_3)_A^{\;B}(\Psi_\mp)_{\nu B}-\frac{1}{2}(\bar{\Psi}_\pm)^A_\mu\gamma^{\mu\nu}\gamma^5 {\rm i} \epsilon e(\sigma_3)_A^{\;B}(\Psi_\mp)_{\nu B}\ ,
       \end{eqnarray}
       and gauge invariance requires
       \begin{equation}
       (\epsilon(y) \Omega(y))'=\frac{e}{g}\epsilon(y)\Longrightarrow \Omega(y)=\frac{e}{g}\epsilon(y)|y|\ ,
       \end{equation}
      hence, one can gauge away non-zero vacuum expectation value by a true gauge transformation.

      One can check, \cite{Bagger:2003fy}, that turning on non-zero gravitini masses on the branes while retaining the ${\bf Z_{2}}$-odd prepotential, explicitly violates the $U(1)$ (gauged and global) symmetry and, therefore, in such a case it becomes possible to break supersymmetry by the expectation value of $ A_5$.

      \section{Summary}
      In this note we have analyzed supersymmetry breakdown by nonzero vacuum expectation value of $A_5$ in five--dimensional warped supergravities on the orbifold ${\bf S^{1}}/{\bf Z_{2}}$. We have shown that typically a wide class of gauge transformations
does not respect the boundary conditions (\ref{warbrzadsbigravit1}), and, equivalently, the  brane actions are not invariant under such transformations. As a consequence, it is not always possible to gauge away a nonzero
vacuum expectation value of  $A_5$. Under such circumstances  supersymmetry
is spontaneously broken and we can parametrize different classes of models by the parameter $\theta\in \langle 0,\frac{1}{g r_c})$,
such that for given gravitini masses on the branes the vacuum expectation value of $A_5$ is $\langle A_5\rangle =\theta+\frac{p}{g r_c}$ for $ p\in {\bf Z}$. In the special
 `tuned' case where $|\lambda_{0}|=|\lambda_{\pi}|=6k$ all gauge transformations respect boundary conditions and supersymmetry remains unbroken for any value of  $\langle A_5\rangle$.
One may say, that the explicit breaking of the $U(1)$ gauge invariance is necessary for supersymmetry breakdown by $\langle A_5\rangle$.
It is interesting to notice that the origin of
supersymmetry violation can be traced back to fermionic boundary conditions
given in terms of boundary mass parameters of gravitini, which are supersymmetry singlets, whereas boundary superpotentials known to play a similar role do
transform under supersymmetry variations.

\vspace*{1em}
When this work was at the final stage of preparation, the very interesting paper \cite{Bagger:2003fy} appeared, where the issue of supersymmetry breakdown by Wilson lines on warped $S^{1}/Z_2$ has been discussed.

\section*{Acknowledgments}
Authors thank M. Quiros and F. Zwirner for interesting conversations.
Z.L. thanks Theory Division at CERN for hospitality.
\noindent This work  was partially supported  by the EC Contract
HPRN-CT-2000-00152 for years 2000-2004, by the Polish State Committee for Scientific Research grants KBN 2P03B 001 25 (Z.L.) and by KBN 2P03B 124 25 (R.M.), and by POLONIUM 2003.

\end{document}